\documentstyle[aps,twocolumn,epsfig]{revtex}

\begin{document}
\draft

\twocolumn[\hsize\textwidth\columnwidth\hsize\csname@twocolumnfalse\endcsname

\title{Thermal and ground-state entanglement in Heisenberg $XX$ qubit rings}
\author{Xiaoguang Wang}
\address{Institute for Scientific Interchange (ISI) Foundation,\\
 Viale Settimio Severo 65, I-10133 Torino, Italy}
\date{\today}
\maketitle

\begin{abstract}
We study the entanglement of thermal and ground states in Heisernberg $XX$ qubit rings with a magnetic field. A general result is found that for even-number rings pairwise entanglement between nearest-neighbor qubits is independent on both  the sign of exchange interaction constants and the sign of magnetic fields. As an example we study the entanglement in the four-qubit model and find that the ground state of this model without magnetic fields is shown to be a four-body maximally entangled state measured by the $N$-tangle. 
\end{abstract}
\pacs{PACS numbers: 03.65.Ud, 75.10.Jm }
]

Quantum entanglement is an important prediction of quantum mechanics and constitutes indeed a valuable resource in quantum
information processing \cite{Bennett}. Much efforts are devoted to the study and characterization of it. Very recently one kind of natural entanglement, the thermal entanglement \cite{Arnesen01,Wang01,Three,PW,WP,Kamta}, is proposed and investigated. The investigations of this type of entanglement, ground-state entanglement \cite{Connor01,Meyer01} in quantum spin models, and relations  \cite{Osborne,Osterloh} between quantum phase transition \cite
{QPT} and entanglement provide a bridge between the quantum information theory and condensed matter physics.

Consider a thermal equilibrium state in a canonical ensemble. In
this situation the system state is described by the Gibb's density operator $%
\rho _T=\exp \left( -H/kT\right) /Z,$ where $Z=$tr$\left[\exp \left(
-H/kT\right) \right] $ is the partition function, $H$ the system
Hamiltonian, $k$ is Boltzmann's constant which we henceforth will take equal
to 1, and $T$ the temperature. As $\rho_T $ represents a thermal state, the
entanglement in the state is called {\em thermal entanglement}\cite
{Arnesen01}. In a recent paper \cite{WP} we showed that in the isotropic Heisenberg $XXX$ model the thermal entanglement is completely determined by the partition function and directly related to the internal energy. In general the entanglement can not be determined only by the partition function. 

In this brief report, we consider a physical Heisenberg $XX$ $N$-qubit ring with a magnetic field and aim to obtain some general results about the thermal and ground-state entanglement. We consider a four-qubit model as an example and examine in detail the properties of entanglement. Both the pairwise and many-body entanglement are considered.

In our model the qubits interact via the following Hamiltonian\cite{Lieb} 
\begin{equation}
H(J,B)=J\sum_{i=1}^N\left( \sigma _{ix}\sigma _{i+1x}+\sigma _{iy}\sigma
_{i+1y}\right) +B\sum_{i=1}^N\sigma _{iz},
\end{equation}
where $\vec{\sigma}_i=(\sigma _{ix},\sigma _{iy},\sigma _{iz})$ is the
vector of Pauli matrices, $J$ is the exchange constant and $B$ is the
magnetic field. The positive and negative $J$ correspond to the
antiferromagnetic (AFM) and ferromagnetic (FM) case, respectively. We assume periodic boundary conditions, i.e., $N+1\equiv 1$. Therefore the model has the symmetry of translational invariance.

It is direct to check that the commutator $[H,S_z]=0,$ (rotation symmetry about the $z$-axis) which guarantees that reduced density matrix $\rho_{12}=\text{tr}_{3,4,\cdots,N}(\rho_T)$ of two nearest-neighbor qubits, say qubit 1 and 2, for
the thermal state $\rho_T$ has the form \cite{Connor01} 
\begin{equation}
\rho _{12}=\left( 
\begin{array}{llll}
u_+ & 0 & 0 & 0 \\ 
0& w & z &0  \\ 
0& z & w & 0 \\ 
0& 0 & 0 & u_-
\end{array}
\right)  \label{eq:rho12}
\end{equation}
in the standard basis $\{|00\rangle ,|01\rangle ,|10\rangle ,|11\rangle \}.$
Here $|ij\rangle\equiv |i\rangle\otimes|j\rangle (i,j=0,1)$, $|0\rangle$ ($|1\rangle$) denotes the state of spin up (down), and $S_{z} =\sum_{i=1}^N\sigma _{iz} /2$ are the collective spin operators. 
The Hamiltonian is invariant under the
transformation $\prod_{n=1}^{N/2}S_{n,N-n+1},$ where $S_{ij}$ is the swap
operator for qubit $i$ and $j$. This is reflection symmetry which implies that the matrix element $z$ is a real number. For any operator $A_{12}$ acting on qubit 1 and 2, we have the relation 
\begin{equation}
\text{tr}_{12}(A_{12}\rho_{12})=\text{tr}_{1,2,\cdots,N}( A_{12}\rho_{T}).
\end{equation} 
Then the reduced density matrix $\rho_{12}$ is directly related to various correlation functions $G_{\alpha \beta }=\langle \sigma$$
_{1\alpha}\sigma_{2\beta}\rangle=$tr$_{1,\cdots,N}(\sigma _{1\alpha }\sigma _{2\beta }\rho _T)$
$(\alpha =x,y,z)$ . Precisely the matrix elements can be written in terms of the correlation functions and the magnetization $M=$tr$(\sum_{i=1}^N\sigma _{iz} \rho_T)$ as 
\begin{eqnarray}
u_+ &=&\text{tr}_{1,\cdots,N}(|00\rangle\langle 00|)
=\frac 14(1+2\bar{M}+G_{zz}),  \nonumber \\
u_- &=&\text{tr}_{1,\cdots,N}(|11\rangle\langle 11|)
=\frac 14(1-2\bar{M}+G_{zz}),  \nonumber \\
z &=&\text{tr}_{1,\cdots,N}(|01\rangle\langle 10|)
=\frac 14(G_{xx}+G_{yy}),  \label{eq:uvz} 
\end{eqnarray}
where $\bar{M}=M/N$ is the magnetization per site. In deriving the above
equation, we have used the translational invariance of the Hamiltonian. 

Due to the fact $[H,S_z]=0$  one has $G_{xx}=G_{yy}$. Then, the concurrence \cite{Con} quantifying the entanglement of two qubits is readily obtained as\cite{Connor01} 
\begin{equation}
C=\max \left[0,|G_{xx}|-\frac 12\sqrt{(1+G_{zz})^2-4\bar{M}^2})\right],  \label{eq:c}
\end{equation}
which is determined by the correlation function $G_{xx}, G_{zz}$, and
the magnetization. In the literature there are a lot of results on 
correlation functions \cite{Corre} in various quantum spin models, and
they may be used for the calculations of entanglement.

By using the translational invariance of the Hamiltonian and the rotational symmetry about the $z$-axis we find an useful relation 
\begin{equation}
G_{xx}=\left( \bar{U}-B\bar{M}\right) /(2J),  \label{eq:relation}
\end{equation}
where $\bar{U}=U/N$ is the internal energy per site and $U$ is the internal
energy. {}From the partition function we can obtain the internal energy and the
magnetization through the well-known relations
\begin{equation}
U=-\frac 1Z\frac{\partial Z}{\partial \beta },M=-\frac 1{Z\beta }\frac{%
\partial Z}{\partial B},  \label{eq:um}
\end{equation}
where $\beta=1/T$. As a combination of Eqs.(\ref{eq:relation}) and (\ref{eq:um}) the correlation function $G_{xx}$ is solely determined by the partition function. 
Therefore the concurrence can be obtained from the partition function and the correlation function $G_{zz}$. We see that the partition function itself is not sufficient for determining the entanglement.

Except for the symmetries used in the above discussions there are also other symmetries in our model. Consider the operator $\Lambda_x \equiv \sigma _{1x}\otimes \sigma _{2x}\otimes \cdot \cdot
\cdot \otimes \sigma _{Nx}$ satisfing $\Lambda_x^2=1$. We have the commutator $[\sigma_{i\alpha }\sigma _{i+1\alpha },\Lambda _x]=0$ and anticommutator $[\sigma
_{iz},\Lambda _x]_{+}=0$. Then we immediately obtain 
\begin{eqnarray*}
G_{\alpha \alpha } &=&\text{tr}\{\Lambda_x\sigma _{1\alpha }\sigma _{2\alpha
}\exp [-\beta H(J,B)] \Lambda _x\}/Z \\
&=&\text{tr}\{\sigma _{1\alpha }\sigma _{2\alpha }\exp [-\beta H(J,-B)]\}/Z,
\\
\bar{M} &=&\text{tr}\{\Lambda_x\sigma _{1z}\exp [-\beta H(J,B)] \Lambda _x\}/Z
\\
&=&-\text{tr}\{\sigma _{1z}\exp [-\beta H(J,-B)]\}/Z.
\end{eqnarray*}
These equations tell us that the correlation function $G_{\alpha \alpha }$
and the square of the magnetization is invariant under the transformation $%
B\rightarrow -B.$ {}From Eq.(\ref{eq:c}) and the invariance of $%
G_{\alpha \alpha }$ and $\bar{M}^2$ we find

{\em Proposition 1. The concurrence is invariant under the transformation }$%
B\rightarrow -B.$

The proposition shows that the pairwise thermal entanglement of the nearest-neighbor qubits is independent on the sign of the magnetic field. And it is valid for both even and odd number of qubits.

In Ref.\cite{Wang01}, we observe an interesting result that the pairwise thermal entanglement for two-qubit $XX$ model with a magnetic field is independent on the sign of the exchange constant $J.$ Now we generalize this result to the case of arbitrary even-number qubits.

For the case of even-number qubits we have
\begin{eqnarray*}
&&H(-J,B)=\Lambda _zH(J,B)\Lambda_z, \\
&&\Lambda_z=\sigma _{1z}\otimes \sigma _{3z}\otimes \cdot \cdot \cdot
\otimes \sigma _{N-1z}.
\end{eqnarray*}
Th transformation $\Lambda_z$ changes the sign of the exchange constant $J$.
Note that the definition of $\Lambda_z$ is different from that of $\Lambda_x$.
It is straightforward to prove  
\begin{eqnarray*}
G_{xx} &=&\text{tr}(\Lambda _z\sigma _{1x}\sigma _{2x}e^{-\beta H(J,B)}\Lambda 
_z) \\
&=&-\text{tr}(\sigma _{1x}\sigma _{2x}e^{-\beta H(-J,B)}), \\
G_{zz} &=&\text{tr}(\Lambda_z\sigma _{1z}\sigma _{2z}e^{-\beta H(J,B)}\Lambda 
_z) \\
&=&\text{tr}(\sigma _{1z}\sigma _{2z}e^{-\beta H(-J,B)}), \\
\bar{M} &=& \text{tr}(\sigma _{1z}e^{-\beta H(-J,B)}).
\end{eqnarray*}
From these equations we see that the absolute value $|G_{xx}|,$ the correlation function $G_{zz},$ and the magnetization are invariant under the transformation $ J \rightarrow -J.$ Hence from Eq.(\ref{eq:c}) and the proposition 1 we arrive at

{\em Proposition 2. For the even-number }$XX${\em \ model with a magnetic
field the pairwise thermal entanglement of nearest-neighbor qubits is independent on the sign of exchange constant $J$ and the sign of magnetic field $B$.}

The proposition shows that the entanglement of AFM qubit rings is the same as that of FM rings. 

Now we consider the case of no magnetic fields. Then the magnetization will
be zero and Eq.(\ref{eq:c}) reduces to $
C=\frac 12\max \left[ 0,|\bar{U}/J|-G_{zz}-1\right]$.
We can prove that the internal energy is always negative. First, {}from the traceless property of the Hamiltonian, it is immediate to check that 
\begin{equation}
\lim_{T\rightarrow \infty}U=
\lim_{T\rightarrow \infty} \frac{\sum_{n}E_n e^{-E_n/T}}{\sum_n e^{-E_n/T}}=0,
\end{equation}
where $E_n$ is the eigenvalue of the Hamiltonian $H$. 
In this limit the correlation function $G_{xx}$ and the magnetization $M$ are also zero. Further from the fact that 
\begin{equation}
\partial U/\partial T=({\langle H^2\rangle-\langle H\rangle^2})/{T^2}={(\Delta H)^2}/{T^2}> 0
\end{equation} 
we conclude that $U$ is always
negative. Then we arrive at

{\em Proposition 3. The concurrence of the nearest-neighbor qubits in the
Heisenberg }$XX${\em \ model without a magnetic field is given by } 
\begin{equation}
C=\left\{ 
\begin{array}{ll}
\frac 12\max [0,-\bar{U}/J-G_{zz}-1] & ~~\text{for AFM,} \\ 
\frac 12\max [0,\bar{U}/J-G_{zz}-1] & ~\text{~for FM.}
\end{array}
\right.  \label{eq:cxxx}
\end{equation}

From the proposition we know that even for the case of no magnetic fields the partition function itself can not determine the entanglement. In the limit of $T\rightarrow 0,$ the above equation reduces to
\begin{equation}
C=\left\{ 
\begin{array}{ll}
\frac 12\max [0,-\frac{E^{(0)}}{NJ}-G_{zz}^{(0)}-1] & ~~\text{for AFM,} \\ 
\frac 12\max [0,\frac{E^{(0)}}{NJ}-G_{zz}^{(0)}-1] & ~\text{~for FM,}
\end{array}
\right. \label{eq:gs}
\end{equation}
where $E^{(0)}$ is the ground-state energy and $G_{zz}^{(0)}$ is the
correlation function on the ground state. Therefore the ground-state
pairwise entanglement of the system is determined by both the ground-state energy and the correlation function $G_{zz}^{(0)}$. Next we consider a four-qubit $XX$ model as an example. 

To study the pairwise entanglement we need to solve the eigenvalue problems of the four-qubit Hamiltonian. We work in the invariant subspace spanned by vectors of fixed number $r$ of reversed spins. The subspace $r=0$ ($r=4$) is trivially containing only one eigenstate $|0000\rangle(|1111\rangle)$ with eigenvalue $4B(-4B)$. The subspace $r=1$ is $4$--dimensional. The corresponding eigenvectors and eigenvalues are given by
\begin{eqnarray}
|k\rangle &=&\frac 12\sum_{n=1}^4\exp (-ink\pi /2)|n\rangle _0(k=0,...,3),
\label{eq:eigen1}
\end{eqnarray}
and $4J\cos (k\pi /2)+2B$, respectively.
Here the `number state' $|n\rangle _0={\cal T}^{n-1}|1\rangle _0$ ,$|1\rangle
_0=|1000\rangle ,$ and ${\cal T}$ is the cyclic right shift operator which commutes
with the Hamiltonian $H$\cite{TTT}. Due to the fact that $H$ commutes $\Lambda _x,$ $|k\rangle {^{\prime }}=\Lambda _x|k\rangle $ is also a eigenstate with
eigenvalue $4J\cos (k\pi /2)-2B.$ The states $|k\rangle$ and $|k\rangle^\prime$ are the so-called W states \cite{Wstate} whose corresponding concurrence is $1/2$. 
We have diagonalized the Hamiltonian in the subspaces of $r=0,1,3,4$. Now we diagonalize the Hamiltonian in the
subspace $r=2$ and use the following notations:
\begin{eqnarray}
|1\rangle _1 &=&|1100\rangle ,|n\rangle _1={\cal T}^{n-1}|1\rangle _1(n=1,2,3,4), 
\nonumber \\
|1\rangle _2 &=&|1010\rangle ,|m\rangle _2={\cal T}^{m-1}|1\rangle _2\text{ }%
(m=1,2).
\end{eqnarray}
The action of the Hamiltonian is then described by
\begin{equation}
H|n\rangle _1=2J\sum_{m=1}^2|m\rangle _2,\; H|m\rangle
_2=2J\sum_{n=1}^4|n\rangle _1.
\end{equation}
By diagonalizing the corresponding 
$6 \times 6$ matrix the eigenvalues and
eigenstates are given by

\begin{eqnarray}
E_{\pm } &=&\pm 4J\sqrt{2},\,|\Psi _{\pm }\rangle =\frac 1{2\sqrt{2}}\left(
\sum_{n=1}^4|n\rangle _1\pm \sqrt{2}\sum_{m=1}^2|m\rangle _2\right), 
\nonumber \\
E_1 &=&0,\,|\Psi _1\rangle =\frac 1{\sqrt{2}}(|1010\rangle -|0101\rangle ), 
\nonumber \\
E_2 &=&0,\,|\Psi _2\rangle =\frac 1{\sqrt{2}}(|1100\rangle -|0011\rangle ),
\label{eq:eigen} \\
E_3 &=&0,\,|\Psi _3\rangle =\frac 1{\sqrt{2}}(|1001\rangle -|0110\rangle ), 
\nonumber \\
E_4 &=&0,\,|\Psi _4\rangle =\frac 12(|1100\rangle +|0011\rangle -|1001\rangle
-|0110\rangle ).  \nonumber
\end{eqnarray}
{}From Eqs.(\ref{eq:eigen1}) and (\ref{eq:eigen}) all the eigenvalues are obtained as $\pm 4B$ $(1),\pm 2B$ $(2)$, $4J\pm 2B$ $(1),-4J\pm 2B$ $(1),$ $\pm 4J%
\sqrt{2}$ $(1),0$ $(4)$, where the numbers in the parenthesis denote the
degeneracy. Then the partition function simply follows as
\begin{eqnarray}
Z &=&4+2\cosh (4\sqrt{2}\beta J)+2\cosh (4\beta B)  \nonumber \\
&&+4[1+\cosh (4\beta J)]\cosh (2\beta B).  \label{eq:zzz}
\end{eqnarray}
{}From Eqs.(\ref{eq:um}), (\ref{eq:eigen1}), (\ref{eq:eigen}) and (\ref{eq:zzz}), we get
the internal engery, magnetization, and correlation function $G_{zz},$

\begin{eqnarray}
-Z\bar{U} &=&2J\sqrt{2}\sinh (4\sqrt{2}\beta J)+2B\sinh (4\beta B) \nonumber\\
&&+2B[1+\cosh (4\beta J)]\sinh (2\beta B) \nonumber\\
&&+4J\sinh (4\beta J)\cosh (2\beta B), \label{eq:aaa} \\
-Z\bar{M} &=&2\sinh (4\beta B)+2[1+\cosh (4\beta J)]\sinh (2\beta B), 
\nonumber \\
ZG_{zz} &=&2\cosh (4\beta B)-\cosh (4\sqrt{2}\beta J)-1.  \nonumber 
\end{eqnarray}
Then according to the relation (\ref{eq:relation}) the correlation function $%
G_{xx}$ is obtained  as
\begin{eqnarray}
G_{xx}Z&=&-\sqrt{2}\sinh (4\sqrt{2}\beta J)\nonumber\\
&-&2\sinh (4\beta J)\cosh(2\beta B). \label{eq:ggg}
\end{eqnarray}
The combination of Eqs.(\ref{eq:aaa}),(\ref{eq:ggg}), and (\ref{eq:c}) gives  the exact expression of the concurrence. It is easy to see that the concurrence is independent on the sign of $J$ and $B$, which is consistent with the general result given by proposition 2 for even-number qubits.

Figure 1 gives a three-dimensional plot of the concurrence again the temperature and magnetic field. The exchange constant $J$ is choosed to be 1 in the following. We observe a threshold temperature $T_c=2.36338$ after which the
entanglement disappears. It is interesting to see that the threshold
temperature is independent on the magnetic field $B$ for our four-qubit
model. For two-qubit $XX$ model the threshold temperature is also
independent on $B$\cite{Wang01}. We further observe that near the zero temperature there exists a dip when the magnetic field increases from zero. The dip is due to the energy level crossing at point $B_{c1}=2(\sqrt{2}-1)=0.82843.$ 
When $B$ increases form $B=2$ near the zero temperature the entanglement disappears quickly since there is another level crossing point $B_{c2}=2$ after which the ground state becomes $|1111\rangle .$ We also see that it is possible to increase entanglement
by increasing the temperature in the range of magnetic field $B>2.$ 

Now we discuss the ground-state entanglement ($T=0$). When $B<B_{c1}$, the
ground state is $|\Psi _{-}\rangle$ with the eigenvalue $E^{(0)}=-4\sqrt{2}J.
$ It is direct to check that  $G_{zz}^{(0)}=-1/2.$ Then according to Eq.(\ref{eq:gs}), the concurrence is obtained as $C=\sqrt{2}/2-1/4$  $=$ $0.45711.$ When $B_{c1}<B<B_{c2}$ the ground state is $|k=2\rangle ^{^{\prime }}$  and the corresponding concurrence is 1/2. When $B>B_{c2}$ the ground state becomes $|1111\rangle $, so there exists no entanglement.

Thus far we have discussed the {\em pairwise entanglement} in both the
ground state and the thermal state. Now we discuss the many-body
entanglement in the ground state.   Recently Coffman {\it et al}~\cite{Cof00}
used concurrence to examine three-qubit systems, and introduced the concept
of the 3--tangle, as a way to quantify the amount of 3--way entanglement in
three-qubit systems. Later Wong and Christensen~\cite{Won01} generalize
3--tangle to even-number $N$--tangle.  The $N$--tangle is defined as
\begin{equation}
\tau _{1,2,\ldots ,N}\equiv |\langle \psi |\sigma_{1y}\otimes\sigma_{2y}\otimes...\otimes\sigma_{Ny}|\psi
^{*}\rangle |^2,  \label{eq:m1}
\end{equation}
with~$|\psi \rangle $ a multiqubit pure state. The $N$-tangle works only
for even number of qubits.

For the ground state   $|\Psi _{-}\rangle $ we have $|\Psi _{-}^{*}\rangle
=|\Psi _{-}\rangle ,$ and 

\begin{equation}
\sigma_{1y}\otimes\sigma_{2y}\otimes\sigma_{3y}\otimes\sigma_{4y}|\Psi _{-}\rangle =|\Psi _{-}\rangle ,
\end{equation}
i.e., the ground state is an eigenstate of the operator $\sigma_{1y}\otimes\sigma_{2y}\otimes\sigma_{3y}\otimes\sigma_{4y}$. Therefore we find $\tau _{1,2,\ldots ,4}=1,$ which means that the
ground state has four-body maximal entanglement. For another two
ground states when varying the magnetic field it is easy to check that $\tau
_{1,2,\ldots ,4}=0,$ which means that the ground state have no genuine four-body entanglement. Note that the ground state $|k=2\rangle {^{\prime }}$ has pairwise entanglement but no four-body entanglement.

In conclusion we have found that the pairwise thermal entanglement of nearest-neighbor qubits is independent on the sign of exchange constants and the sign of magnetic fields in the $XX$ even-number qubit ring with a magnetic field. For determining the concurrence we need to know not only the partition function but also one correlation function $G_{zz}$. For the four-qubit model we observe that there exists a threshold temperature which is independent on the magnetic field. The effects of level crossing on the thermal entanglement and ground-state entanglement are also discussed. Finally we find that the ground state is a  four-body maximally entangled state according to the potential many-body entanglement measure, $N$-tangle.

\acknowledgments
The authors thanks for the helpful discussions with Paolo Zanardi, Irene
D'Amico, Hongchen Fu, Prof. Allan I. Solomon, Prof. Joachim Stolze, Prof. Guenter Mahler, and  Prof. Vladimir Korepin. This work has been supported by the European Community through grant IST-1999-10596 (Q-ACTA).

\begin{figure}
\begin{center}
\epsfig{width=10cm,file=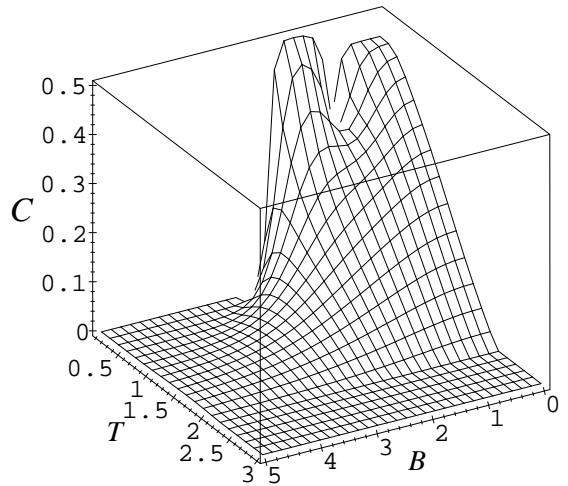}
\caption[]{The concurrence against the temperature and magnetic field.
The parameter $J=1$.} 
\end{center}
\end{figure}

\end{document}